\documentclass[aps,twocolumn,prb]{revtex4}

\usepackage{graphicx}
\usepackage{amssymb}
\usepackage{amsmath}

\newcommand{\R}{\mathbf{r}}
\newcommand{\up}{\uparrow}
\newcommand{\dw}{\downarrow}

\begin{document}

\title{Generalized gradient approximation correlation energy functionals based on the uniform 
electron gas with gap model}

\author{Eduardo Fabiano}
\email{eduardo.fabiano@nano.cnr.it}
\affiliation{National Nanotechnology Laboratory (NNL), Istituto di Nanoscienze-CNR, Via per Arnesano 16, I-73100 Lecce, Italy }
\author{Paolo E. Trevisanutto}
\affiliation{Center for Biomolecular Nanotechnologies @UNILE, Istituto Italiano di Tecnologia, Via Barsanti, I-73010 Arnesano, Italy}
\author{Aleksandrs Terentjevs}
\affiliation{National Nanotechnology Laboratory (NNL), Istituto di Nanoscienze-CNR, Via per Arnesano 16, I-73100 Lecce, Italy }
\author{Lucian A. Constantin}
\affiliation{Center for Biomolecular Nanotechnologies @UNILE, Istituto Italiano di Tecnologia, Via Barsanti, I-73010 Arnesano, Italy}

\begin{abstract}
We study the uniform electron gas with a gap model in the context of density functional
theory. Based on this analysis, we construct two local gap models that realize
generalized gradient approximation (GGA) correlation
functionals satisfying numerous exact constraints for the correlation energy.
The first one, named GAPc, fulfills 
the full second-order correlation gradient expansion at any density regime,
is very accurate for jellium surfaces, comparable to
state-of-the-art GGAs for atomic systems and molecular systems, and well 
compatible with known semilocal exchange. 
The second functional, named GAPloc, is satisfying 
the same exact conditions,
except that the second-order gradient expansion is sacrificed for a better 
behavior under the Thomas-Fermi scaling and a more realistic correlation 
energy density of the hellium atom. The GAPloc functional
displays a high accuracy for atomic correlation energies, still preserving
a reasonable behavior for jellium surfaces. Moreover, it shows 
a higher compatibility
with the Hartree-Fock exchange
than other semilocal correlation functionals. 
This feature is explained in terms of the real-space analysis of the GAPloc correlation energy. 
\end{abstract}

\maketitle

\clearpage

\section{Introduction}
Model systems play a prominent role in Kohn-Sham (KS) \cite{ks}
density functional theory (DFT) \cite{parrbook}, due to their utility in
understanding and guiding the development
of physically sound and accurate exchange-correlation (XC) functionals,
which shall include quantum effects of the many-electron interaction.
Thus, over the years, a wide number of model systems has been
considered within DFT \cite{ks,PhysRevLett.81.3487,
PhysRevB.62.10046,arpa,PhysRevB.79.115117,PhysRevB.56.3655,
PhysRevB.1.4555,PhysRevLett.100.256406,3059783,apbe,
0953-8984-12-7-308,PhysRevB.78.155106,q2dgga,466658,pbeloc,
PhysRevB.82.115103,ldagap,PhysRevB.87.205143,pittalis08,aa3675681,Tognetti2008536}.

The first, and probably most important, model system of DFT has been
the uniform electron gas \cite{ks} that is a paradigm for
solid-state physics and most non-empirical XC functionals
\cite{scusrev}. The uniform electron gas model leads straightforwardly
to relatively simple explicit expressions for both
the exchange and the correlation energy of a many electron system,
within the so called local density approximation (LDA)
\cite{ks,dirac29,PhysRev.81.385,vwn,pwlda}. This approximation, despite its simplicity, 
provides a remarkably good qualitative description of
inhomogeneous electronic systems, however, it fails severely
to yield accurate quantitative results.

For this reason many improvements over the LDA level of theory have been
proposed \cite{scusrev}. In particular, by using the gradient of the density as an
additional input ingredient, useful generalized 
gradient approximation \cite{gga} (GGA) functionals have been developed,
which provide a good compromise between simplicity,
computational efficiency, and numerical accuracy
\cite{arpa,apbe,pbe,revpbe,b88,lyp,pbeint,zpbesol,zvpbesol,pbeintgold,am05_1,am05_2,pittalis09,MdelCampo2012179,3701132,3691197,Tognetti2007381,tognetti10,QUA22571,bremond12,bremond13}.
Alternatively, for the correlation energy, Rey and Savin \cite{ldagap}
proposed to consider a model system (jellium with gap)
obtained by adding a non-local
one-body operator to the true Hamiltonian of the uniform electron
gas, in order to force an arbitrary gap $G$ between occupied and virtual
orbitals. In this way, in fact, the strong overestimation of the
LDA correlation can be largely corrected.
Moreover, for the jellium with gap model system the correlation energy can be
computed with high accuracy for different values of $G$. Then,
an analytic representation of the LDA with gap correlation
functional is readily available \cite{ldagap}.

This formula can be easily extended to be used for inhomogeneous systems after
providing, in analogy with the usual LDA approach, an appropriate
local description of the gap (i.e. a function $G[n,\nabla n,\ldots](\R)$).
This idea was used as a seed to construct the
KCIS \cite{kcis} and KCISK \cite{kcisk} 
correlation functionals which showed, with their
good performance, the advantages of the jellium with gap model system.
However, in the construction of the KCIS/KCISK functional 
only a limited focus was posed on the local gap functional and a
rather simple ansatz was used for the gap $G$. In contrast, 
several important constraints of the exact correlation energy,
were satisfied using the conventional approach of GGA and meta-GGA
functionals. In particular, the LDA for the uniform electron gas at $G=0$ 
was substituted by a GGA expression to assure the recovery of the
second-order gradient expansion for the correlation energy (only at $G=0$)
\cite{kcis,kcisk}. 

In this paper we want to resume the work on the jellium with
gap correlation and explore the possibility
to construct appropriate local gap models in order to 
achieve an accurate description of inhomogeneous electronic systems.
In this sense, our work differs from the one performed for the KCIS/KCISK
functionals \cite{kcis,kcisk}, 
because of the stronger attention we put on the construction
of the local gap function. We show in fact that it is possible,
in general, to recover many exact properties of the correlation energy
by building a suitable local gap function $G$, being a functional of 
the density and its gradient. 
The focus of our work will be on the GGA level of theory because,
being this the most simple beyond LDA, it is also the most powerful
for showing the significance of various exact conditions and for exploring and explaining 
the physics related to a particular model system. Thus, by developing two, conceptually 
different, local gap models, we aim at showing the power of this approach.
At the same time, our functionals may be also considered the basis
for more sophisticated developments, even beyond the GGA level.
Nevertheless, in this paper we address only the GGA level and thus all 
our results are compared with the GGA state-of-the-art 
ones.

\section{Theory} 
Within the jellium with gap model \cite{ldagap,kcisk} the correlation energy is defined as
\begin{equation}\label{ee1}
E_c = \int n(\R)\epsilon_c(r_s,\zeta,G)d\R \ ,
\end{equation}
where $n$ is the electron density, $r_s=(3/4\pi n)^{1/3}$ is the local 
Seitz radius, $\zeta=(n_\up-n_\dw)/n$ is the relative spin polarization,
$G$ is the gap, and
\begin{equation}\label{e2}
\epsilon_c(r_s,\zeta,G) = \epsilon_0(r_s,G) + f(\zeta)\left[\epsilon_1(r_s,G)-\epsilon_0(r_s,G)\right]\ .
\end{equation}
In Eq. (\ref{e2}) the spin mixing factor is
\begin{equation}
f(\zeta) = \frac{(1+\zeta)^{4/3} + (1-\zeta)^{4/3} - 2}{2^{4/3}-2}\ ,
\end{equation}
while the fully-spin-polarized (${\epsilon}_1$)
and the spin-unpolarized ($\epsilon_0$) correlation energies
per particle have the form (the spin index is omitted hereafter for notational
simplicity)
\begin{equation}\label{ee4}
\epsilon(r_s,G) = \frac{\epsilon_c^{LDA}(r_s) + c_1(r_s)G}{1+c_2(r_s)G+c_3(r_s)G^2}\ ,
\end{equation}
with $\epsilon_c^{LDA}$ being the conventional 
local density approximation for the correlation energy and
$c_1$, $c_2$, and $c_3$ some density-dependent coefficients \cite{ldagap}.
Full details about the construction of the model as well as its asymptotic
properties are discussed in the appendix.

The main goal of this work is finding appropriate 
local gap functions $G(r_s,t)$ (with $t=|\nabla n|/(4\phi(3/\pi)^{1/6}n^{7/6})$ 
and $\phi=((1+\zeta)^{2/3}+(1-\zeta)^{2/3})/2$), such
that exact contraints of the correlation energy can be imposed
to the jellium with gap model. The idea behind this 
approach is that the LDA model with a gap can
accurately describe, at each point in space, 
the correlation energy per particle
of an inhomogeneous electron distribution if
a proper gap is applied at each point,
depending on the local/semilocal properties of the
electron densities at that point. 

To this end, we consider in the appendix a
detailed analysis of several known exact constraints for the
correlation energy and their relation with
local gap function. In particular we consider the
uniform electron gas limit ($\nabla n = 0$), the
slowly-varying density limit ($t\rightarrow 0$), the
rapidly-varying density limit ($t\rightarrow\infty$),
the uniform scaling to the high-density limit \cite{PhysRevA.45.1509},
and the Thomas-Fermi scaling \cite{PhysRevLett.100.256406,scaling}.
Using this information we can propose two local gap functions
able to enforce a set of exact constraints on the jellium with
gap correlation.

With the aim of imposing the second-order gradient expansion behavior
we consider the general formula of Eq. (\ref{e25}) and set
\begin{equation}
G_{GAPc}(r_s,t^2,\zeta) = \phi^3 \frac{\beta(r_s)t^2}{c_1-c_2\epsilon_c^{LDA}}H(r_s,t^2)\ ,
\end{equation}
with
\begin{equation}
H(r_s,t^2) = \frac{a+ \left[Ar_s\log(r_s)\beta^{-1}(r_s)\right]t^2}{a + t^2}\ ,
\end{equation}
where $\beta$ is the $r_s$-dependent second-order correlation
coefficient of Hu and Langreth \cite{PhysRevB.33.943} (we use the parametrization 
of Ref. \citenum{revtpss}) and $a=30$ is a parameter fixed by minimizing 
the variance of the correlation energy error for the noble gas atoms He, Ne, 
and Ar. The resulting functional is labeled GAPc and satisfies all the 
exact constraints mentioned before (i.e. the uniform
electron gas and the slowly-varying limits, the rapidly-varying
density limit, the uniform scaling towards the high-density limit,
and the Thomas-Fermi scaling behavior; see the appendix for details).
Thus, it formally fulfills the same exact conditions as the PBE correlation
functional \cite{pbe}. However, the GAPc functional
recovers the true second-order gradient expansion at any $r_s$,
whereas PBE does it only in the high-density limit.
Moreover, in the tail of an atomic density
$G_{GAPc}\rightarrow\lambda\log(\lambda)$,
with $\lambda\propto r_s \rightarrow\infty$,
(note that $c_1-c_2\epsilon_c^{LDA}\rightarrow const$), 
therefore the GAPc functional has the tail decay 
$\epsilon_c^{GAPc}\propto 1/(\lambda^4\log(\lambda))$,
closer
 to the behavior described in Ref. \citenum{PhysRevB.31.3231}
(note that $\log(\lambda)\propto r$). 

A second useful local gap function can be obtained by dropping the
requirement that the second-order gradient expansion behavior
is satisfied. This condition was shown in fact to
be not very important for many cases \cite{PhysRevB.79.201106}.
In this way the choice of the local gap function is no more
bound to Eq. (\ref{e25}) and a more flexible ansatz can be
employed. Our choice, which defines the GAPloc functional
is
\begin{equation}
G_{GAPloc}(r_s,s,t) = f_G\frac{s^{\alpha(t)+2}}{r_s^2}\frac{b+s^2}{1+s^{2+\alpha(t)}}\ ,
\end{equation}
where $s=|\nabla n|/(4(3\pi^2)^{2/3}n^{4/3})$, $f_G=1/(128\pi^22^{2/3})$, and
\begin{equation}
\alpha(t) = \frac{\alpha_1+t^3}{1+t^3}\ .
\end{equation}
Note that $f_Gs^2/r_s^2=\tau^W/n$ ($\tau^W$ being the 
von Weizs\"aker kinetic energy density \cite{vw}) is the
local gap function used in the KCIS/KCISK functionals \cite{kcis,kcisk}.
The two parameters $b=14.709$ and $\alpha_1=6.546$
were fixed by fitting to the exact correlation
energy per particle of the He atom 
(i.e. the function $\epsilon_c^{He}$ reported
in Ref. \citenum{Barendes1}).
We recall that fitting to the energy density of model
systems is a common practice in DFT, used for instance 
in the construction of the AM05 \cite{am05_1,am05_2} 
and ARPA+ \cite{arpa} GGA functionals, which 
may help to reduce the error cancellation effects
in contrast to fitting to integrated energies. 
The GAPloc functional satisfies the following exact constraints:

(1) LDA limit; in fact for $|\nabla n|=0$ we have $G_{GAPloc}=0$.
Note however that for $|\nabla n|\rightarrow 0$ the functional behaves
as
\begin{equation}
\epsilon_c^{GAPloc} \propto\epsilon_c^{LDA} + 
f_Gb\frac{s^{2+\alpha_1}}{r_s^2}\left(c_1-c_2\epsilon_c^{LDA}\right)\sim \epsilon_c^{LDA}\ .
\end{equation}
Thus, as anticipated, the GAPloc functional does not recover the
second-order gradient expansion behavior but approaches the
LDA limit much faster.

(2) Rapidly-varying density limit; for $|\nabla n|\rightarrow \infty$ and $r_s$ finite,
$G_{GAPloc} \propto s^2/r_s^2 \rightarrow\infty$ and the correlation correctly vanishes.

(3) Uniform scaling to the high density-limit; in this limit
$G_{GAPloc}\propto\lambda^2$. Thus, according to Eq. (\ref{e26}) 
$\epsilon_c^{GAPloc}\rightarrow-f_C$ and the logarithmic divergence
of LDA is canceled. 

(4) Thomas-Fermi scaling; in fact in this limit we have
\begin{equation}
G_{GAPloc}\propto\lambda^\frac{2-\alpha_1+t^3}{3(1+t^3)}\sim \lambda^\frac{-4.5+t^3}{3(1+t^3)}
\end{equation}
and $(-4.5+t^3)/(3(1+t^3))<1$ for any positive $t$.

The tail behavior of the GAPloc functional can be also easily 
inspected noting that in the tail of an atomic density
$s^2\propto \lambda^2$, with $\lambda\propto r_s\rightarrow\infty$ (see appendix). 
Thus, the local gap function behaves as
\begin{equation}\label{e42}
G_{GAPloc} \propto -\epsilon_H\frac{\lambda^{\alpha(\lambda)+2}}
{\lambda^2}\frac{b+\lambda^2}{1+\lambda^{\alpha(\lambda)+2}}
\sim -\epsilon_H\left(1+\frac{b}{\lambda^2}\right)\ ,
\end{equation}
where $\alpha(\lambda)$ is weakly dependent on $\lambda$.
Equation (\ref{e42}) indicates that as $\lambda\rightarrow\infty$ then
$G_{GAPloc}\rightarrow-\epsilon_H$ therefore the GAPloc functional (a.i. $\epsilon_c^{GAPloc}$)
decays as $1/\lambda^2$. However, because the actual value of the $b$
parameter is rather large this asymptotic behavior is set
only quite far in the tail, whereas at smaller distances the
local gap function is quite bigger. As a consequence the
GAPloc correlation energy per particle will
decrease rather fast in the outer valence region and then
attains a $1/\lambda^2$ decay behavior in the far tail.

\begin{figure}
\begin{center}
\includegraphics[width=\columnwidth]{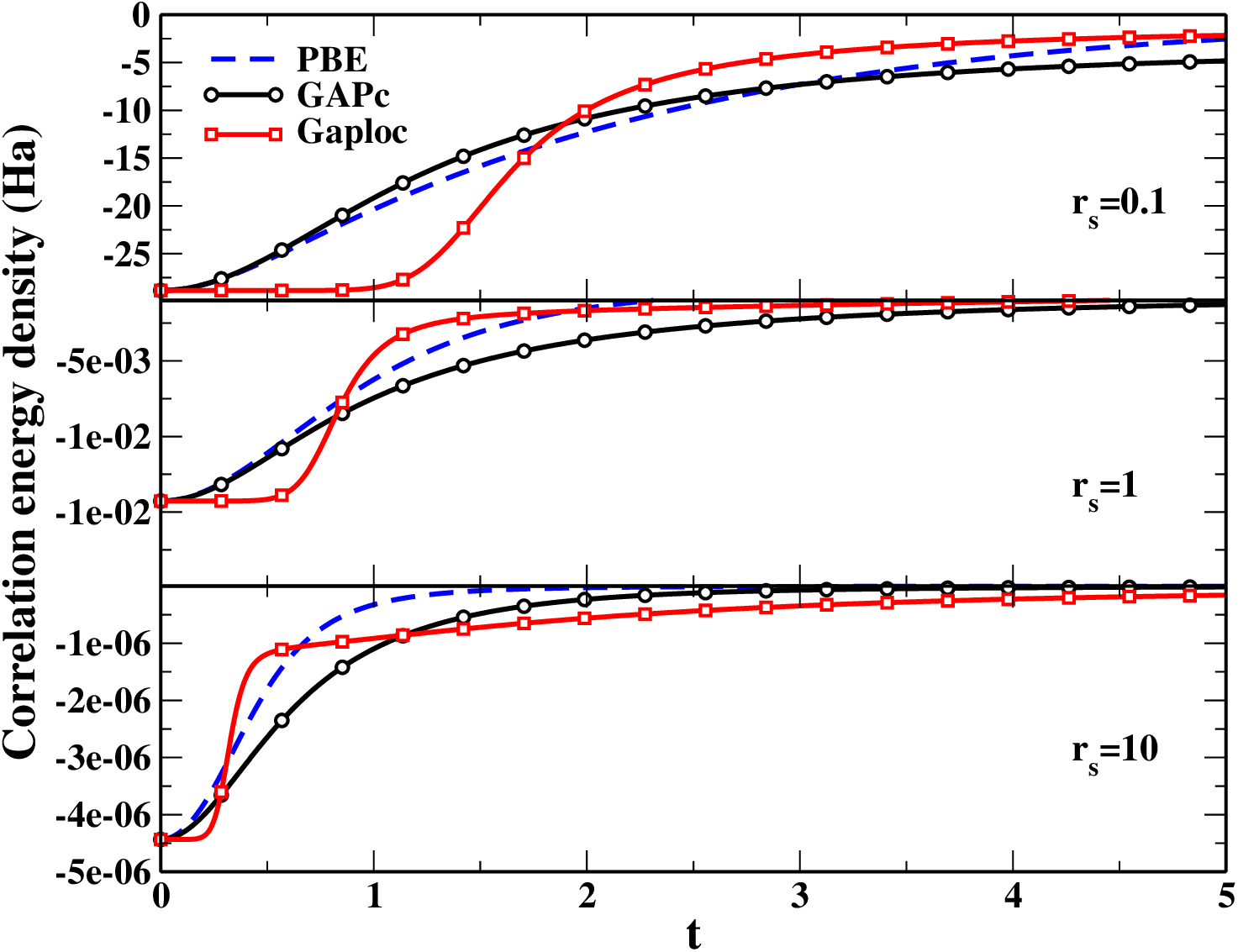}
\caption{\label{fig1}Correlation energy density of different functionals as a function of the 
reduced gradient $t$ for several values of $r_s$ (spin-unpolarized case; $\zeta=0$).}
\end{center}
\end{figure}

The different features of the GAPc and GAPloc functionals that we
discussed above are summarized in Figure \ref{fig1} where 
we report a plot of the correlation energy densities for
several values of $t$ and $r_s$ and compare them with the
the PBE functional.
Similar plots are obtained for the KCIS/KCISK correlation energy densities.
The figure highlights in particular the similarity
of the GAPc and PBE functionals for small $t$ and small $r_s$,
because both functionals recover the same second-order gradient
expansion in this high-density limit. The similarity is 
much less for larger values of the $r_s$ Seitz parameter, because in
this case the PBE functional fails to recover the correct
value of $\beta$.  Another important feature that
emerges from the plot is the peculiar shape of the 
GAPloc functional. In fact, as discussed above,
this functional recovers the LDA limit for
a much broader interval of small $t$ values but then
shows a sudden reduction of the correlation energy density
at $t\approx1$ before starting a slow decay.
This behavior corresponds, despite the slow asymptotic
decay behavior, to a localization of the correlation energy density
near the atomic nuclei. This is an important feature
of this functional that will be discussed in more details later.

\begin{figure}
\begin{center}
\includegraphics[width=\columnwidth]{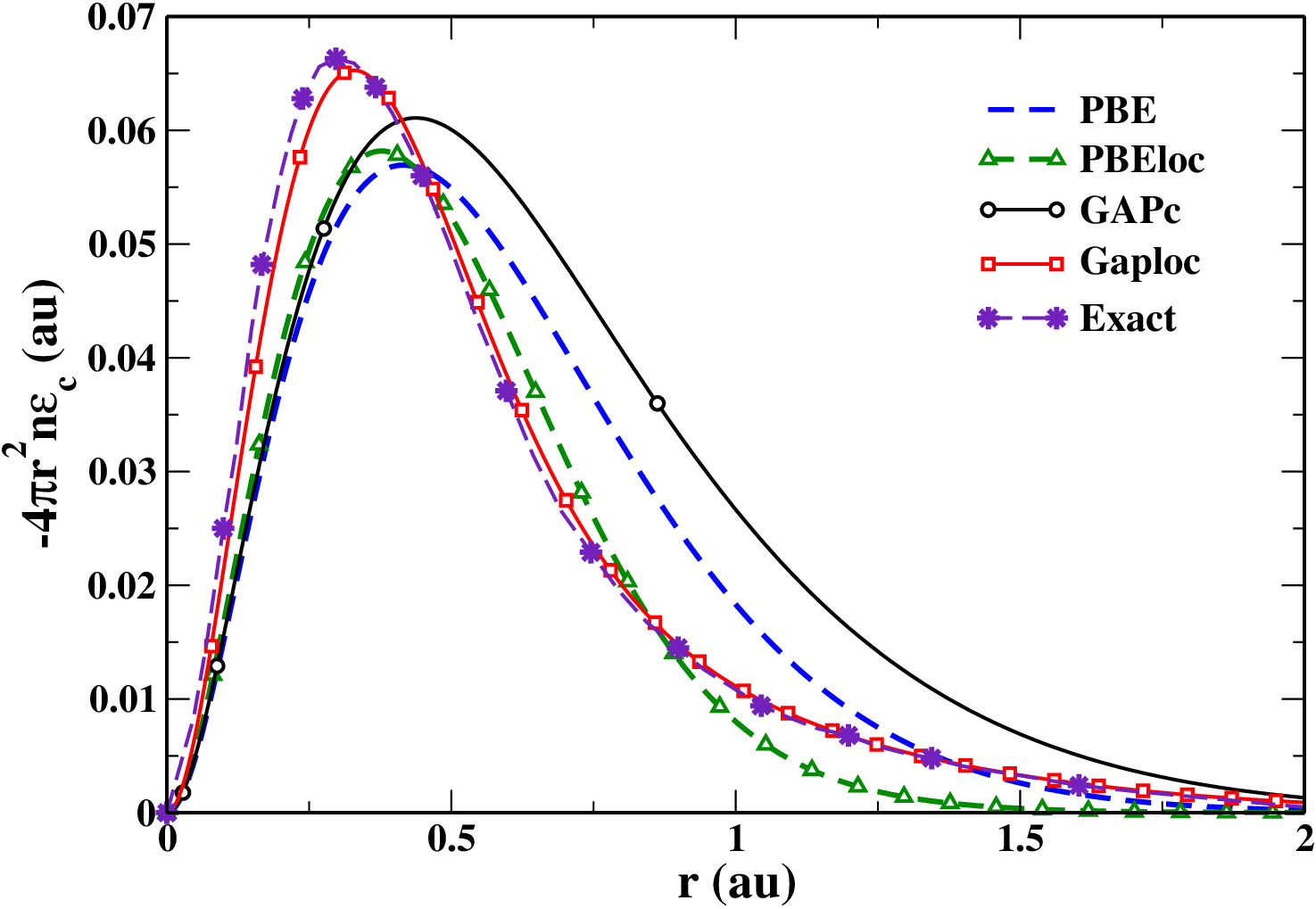}
\caption{\label{fig1b}
$-4\pi r^2 n\epsilon_c$ versus the radial distance $r$ for
the He atom. The exact curve is from Ref. \citenum{Barendes1}.
}
\end{center}
\end{figure}
Finally, we show in Fig. \ref{fig1b}, the opposite of the radial 
correlation energy density ($-4\pi r^2 n\epsilon_c$) 
versus the radial distance $r$ for the He atom. 
By construction, GAPloc is very close to the exact curve. 
Note instead that a functional that recovers the 
second-order gradient expansion in general can 
not be accurate near the nucleus, giving a too small 
(in absolute value) correlation energy density at small $r$. 
Then, assuming that this functional is accurate for the 
total correlation energy ($E_c$), it must give a 
too large (in absolute value) correlation energy density at 
larger $r$. Hence, such a correlation energy density is 
quite different from the exact behavior, being modified up 
to a gauge transformation \cite{gauge1}. 
Such a gauge had been the subject of intense 
research, at the hyper-GGA level of theory \cite{hyper1}.
Both PBE and GAPc show this semilocal-gauge behavior, 
while PBEloc \cite{pbeloc} is definitely better due to 
the localization concept which incorporates \cite{pbeloc}. 
Note also that GAPc is slightly worse than PBE (and very similar to revTPSS instead), 
because it recovers the full (correct) second-order gradient expansion and
its total correlation energy ($E_c$) is slightly worse (see Table \ref{tab2}).

\section{Computational details}
To test the correlation functionals we performed several calculations
on different test sets that are summarized below.
We remark that the selection of tests for the assessment of the
GAPc functional in conjunction with the revPBE exchange \cite{revpbe} was
restricted to those properties and systems where the revPBE exchange
may be expected to yield reasonable results \cite{mukappa} (e.g. atomization energies
of organic molecules, atomic properties). We did not consider instead tests
where the revPBE exchange is completely inadequate (e.g. structural properties,
metallic systems, solids), because in these cases it would be 
impossible to extract any useful information on the performance
of the correlation, which is the target of the present study,
due to the dominating exchange error.

The following sets of properties were considered
\begin{itemize}
\item \textbf{Atomic and ionic correlation energies}. The correlation energies of a set of
24 atoms and ions were used as benchmark \cite{js,PhysRevA.44.7071,QUA:QUA2}. In addition,
we considered the second-order M\o ller-Plesset estimates of Ref. \citenum{3547262}
as well as the results of virial-constrained effective Hamiltonian method
of Ref. \citenum{QUA:QUA2}.
\item \textbf{Jellium surface correlation energies}. We compared the
surface correlation energies of semi-infinite jellium 
surfaces with bulk parameter $r_s=2$, 3, 4, and 6. Reference data were taken
from diffusion Monte Carlo (DMC) calculations \cite{PhysRevB.76.035403}.
\item \textbf{Atomization energies}. We assessed the atomization energies of organic
molecules from the AE6 \cite{ae6} and the W4 \cite{w4} test sets; in addition, for
transition metal complexes we considered the TM10AE set \cite{zpbesol}.
\item \textbf{Organic reactions}. Reaction energies and barrier heights
from the BH6 \cite{ae6}, K9 \cite{k9}, and DC9 \cite{doi:10.1021/ct3002656} test sets
have been calculated.
\item \textbf{Other properties} We considered the G21P set of 
ionization potentials of atoms and molecules \cite{doi:10.1021/ct900489g,C0CP02984J},
the EA13 test of electron affinities of atoms and molecules \cite{doi:10.1021/jp021590l},
the PA13 set comprising proton affinities of organic molecules \cite{doi:10.1021/ct900489g},
and the AE17 test set of atomic non-relativistic exchange-correlation energies \cite{m06}. 
In addition, we tested the HB6 set of hydrogen-bond complexes \cite{doi:10.1021/ct049851d}.
\end{itemize}

Molecular calculations were carried out self-consistently with the 
TURBOMOLE program package \cite{turbomole} using the
def2-TZVPP basis set \cite{1627293,B508541A}.
Atomic calculations for Table \ref{tab2} were performed
using a cc-pV5Z basis set \cite{466439,464303,456153,1998907}.
Atomic calculations for Fig. \ref{fig2} were carried out with the
Engel code \cite{engelcode}, using accurate exact exchange orbitals and densities. 
Jellium calculations were performed with numerical Kohn-Sham LDA
orbitals and densities. FORTRAN90 routines implementing both 
functionals are freely available on the web \cite{subroutines}.

\section{Numerical results}
In this section we report the results of some numerical calculations that
we performed to assess the performance of the correlation functionals
and study their properties in different contexts.

\subsection{Atomic and jellium correlation energies}
Table \ref{tab2} reports the correlation energy per electron of several
atoms and ions as computed with different correlation functionals.
\begin{table*}
\begin{center}
\caption{\label{tab2} Correlation energy (mHa) divided by the number of electrons ($N_e$) for several atoms and ions. Reference data are taken from Refs.\citenum{3547262} and \citenum{QUA:QUA2}. The last lines report the mean error (ME), the mean absolute error (MAE), the mean absolute relative error (MARE), and the standard deviation for each series. The best result of each line is denoted in bold face.}
\begin{ruledtabular}
\begin{tabular}{lrccccccc}
System  &  $N_e$  &  LDA  &  LYP  &  PBE  &  APBE  &  GAPc   &  GAPloc  & Ref.  \\ 
\hline
He  &  2  &  -56.2  &  -21.9  &  \textbf{-21.0}  &  -18.7  &  -26.2  &  -20.0  & -21.0  \\
Li$^+$  &  2  &  -67.3  &  -23.8  &  \textbf{-22.4}  &  -19.8  &  -27.6 &  -20.4  & -21.7  \\
Be$^{2+}$  &  2  &  -75.2  &  -24.5  &  \textbf{-23.0}  &  -20.3  &  -28.0  &  -20.0  & -22.2  \\
Li  &  3  &  -50.3  &  -17.8  &  -17.1  &  \textbf{-15.2}  &  -21.4 &  -15.9  & -15.1  \\
Be$^+$  &  3  &  -57.6  &  -20.4  &  -18.1  &  \textbf{-15.9}  &  -22.3  &  -16.1  & -15.8  \\
B$^{2+}$  &  3  &  -63.2  &  -22.3  &  -18.6  &  \textbf{-16.3}  &  -22.7  &  -15.8  & -16.2  \\
C$^{3+}$  &  3  &  -67.7  &  -23.7  &  -18.9  &  \textbf{-16.6}  &  -22.9  &  -15.5  & -16.5  \\
N$^{4+}$  &  3  &  -71.5  &  -24.8  &  -19.1  &  \textbf{-16.7}  &  -22.9  &  -15.2  & -16.7  \\
O$^{5+}$  &  3  &  -74.9  &  -25.6  &  -19.2  &  \textbf{-16.8}  &  -22.8  &  -14.9  & -16.8  \\
Ne$^{7+}$  &  3  &  -80.4  &  -26.9  &  -19.4  & \textbf{ -16.9}  &  -22.7  &  -14.4  & -17.0  \\ 
Ar$^{15+}$  &  3  &  -94.9  &  -29.2  &  -19.7  &  \textbf{-17.1}  &  -22.2  &  -13.9  & -17.4  \\ 
Be  &  4  &  -56.0  &  \textbf{-23.6}  &  -21.4  &  -19.3  &  -25.7  &  -20.2  & -23.6  \\
B$^+$  &  4  &  -63.0  &  -26.7  &  -23.0  &  -20.8  & \textbf{-27.4}  &  -21.7  & -27.8  \\ 
C$^{2+}$  &  4  &  -68.5  &  \textbf{-28.6}  &  -24.0  &  -21.6  &  -28.4  &  -22.3  & -35.1  \\ 
N$^{3+}$  &  4  &  -73.0  &  \textbf{-30.0}  &  -24.7  &  -22.2  &  -29.0  &  -23.0  & -35.1  \\ 
O$^{4+}$  &  4  &  -76.9  &  \textbf{-30.9}  &  -25.3  &  -22.7  &  -29.4  &  -23.7  & -38.5  \\ 
N  &  7  &  -61.0  &  \textbf{-27.4}  &  -25.7  &  -23.4  &  -28.8  &  -25.8  & -26.9  \\ 
O$^+$  &  7  &  -65.6  &  -29.5  &  \textbf{-27.0}  &  -24.5  &  -29.9  &  \textbf{-27.0}  & -27.7  \\ 
Ne  &  10  &  -74.3  &  -38.4  &  -35.1  &  -32.3  &  -38.2  &  \textbf{-38.5}  & -39.1  \\ 
Ar$^{8+}$  &  10  &  -96.8  &  -44.9  &  \textbf{-41.0}  &  -37.6  &  -42.0  &  -45.0  & -39.9  \\ 
Ar$^{6+}$  &  12  &  -90.2  &  -44.8  &  -38.3  &  -35.0  &  -40.1  & \textbf{-40.8}  & -41.3  \\ 
Ar  &  18  &  -79.1  &  -41.7  &  \textbf{-39.3}  &  -36.4  &  -41.0  &  -43.0  & -40.1  \\ 
Zn  &  30  &  -88.5  &  -47.7  &  -46.9  &  -43.6  &  -47.3  &  \textbf{-52.6}  & -56.2  \\ 
Kr  &  36  &  -90.8  &  -48.6  &  -49.1  &  -45.8  &  -48.8  &  \textbf{-56.1}  & -57.4  \\ 
  &    &    &    &    &    &    &    &   \\ 
ME  &&  -44.1  &  -1.6  &  2.0  &  4.6  &  \textbf{-1.4} &  2.6     &   \\ 
MAE  &&  44.1  &  4.8  &  3.8  &  4.6  &  4.9  &  \textbf{3.4}      &   \\ 
MARE  &&  192\%  &  21\%  &  13\%  &  13\%  &  21\%  &  \textbf{11\%}      &   \\ 
St. Dev. & &  11.8  &  5.8  &  5.0  &  5.1  &  5.5  &  \textbf{4.7}  &       \\
\end{tabular}
\end{ruledtabular}
\end{center}
\end{table*}
All the GGA functionals reported in the table perform 
rather similarly and improve considerably over the LDA functional.
Nevertheless, the best results are obtained by the GAPloc functional which
yields a mean absolute error (MAE) of 3.4 mHa and
a mean absolute relative error of 11\%. Slightly larger errors are
yielded by the PBE functional. All other GGAs here are roughly 1 mHa 
worst on average. 
In particular, the GAPc functional gives a MAE of 4.9 mHa,
slightly better than the LYP functional.
Note also that KCIS performs similarly to PBE
(MAE 3.8 kcal/mol), whereas KSICK yields results comparable to
GAPloc (MAE 3.3 kcal/mol).

\begin{figure}
\begin{center}
\includegraphics[width=\columnwidth]{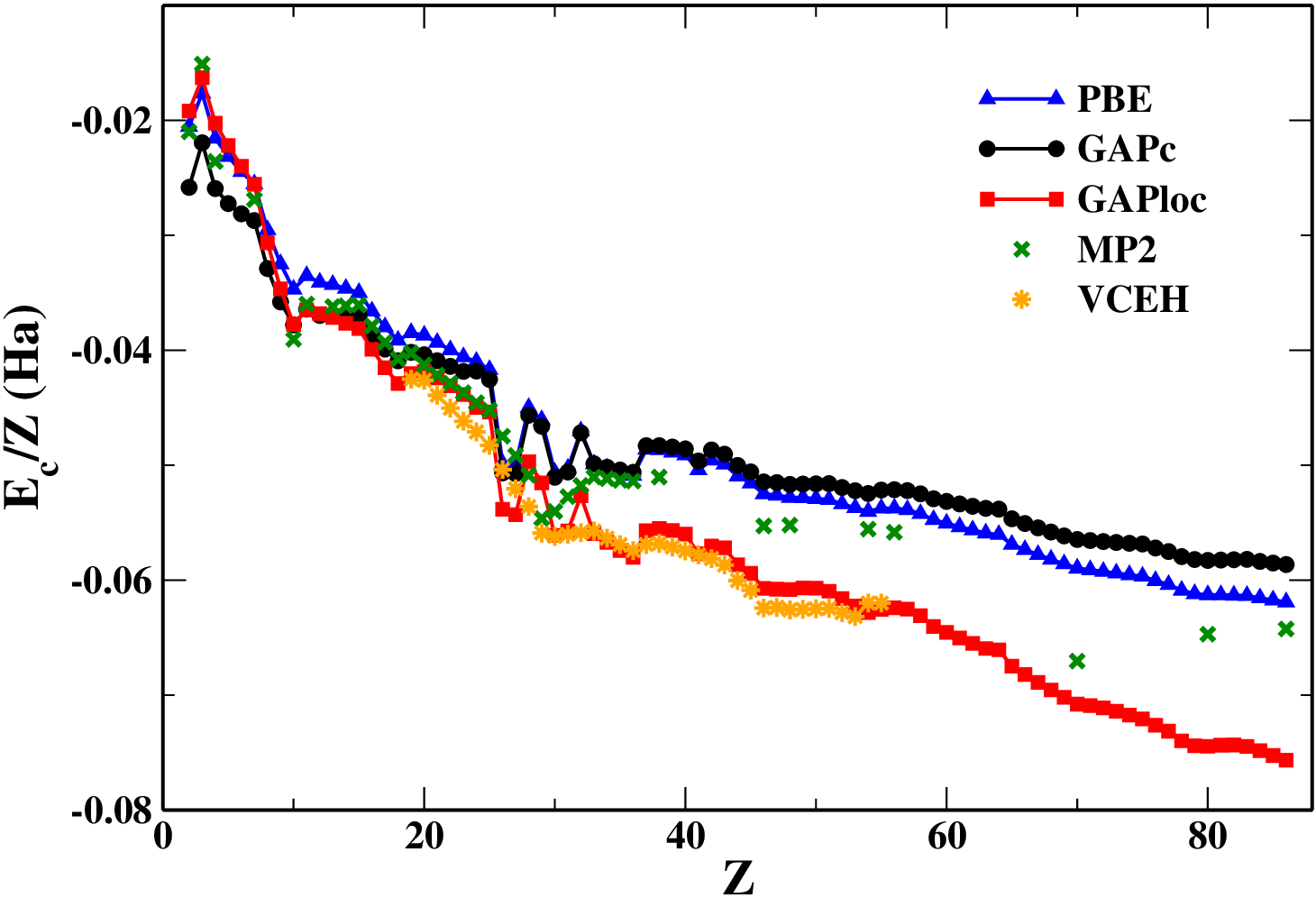}
\caption{\label{fig2}Correlation energy per electron of different atoms versus the atomic number 
$Z$. As a reference also MP2 \cite{3547262} and virial-constrained effective Hamiltonian (VCEH) method 
\cite{QUA:QUA2} results are reported.}
\end{center}
\end{figure}

These results are extrapolated to larger atoms in Fig. \ref{fig2} where we
report the correlation energy per electron of atoms up to
Fr ($Z=87$) as computed with different functionals.
We remark that in this case the comparison is only semi-quantitative,
because highly accurate benchmark results are missing for heavier atoms. 
Nevertheless, good reference data are obtained by the virial-constrained
effective Hamiltonian (VCEH) method \cite{QUA:QUA2}.
The plot confirms the remarkable accuracy of the GAPloc functional
even for heavy atoms. At the same time it shows that the
GAPc functional is comparable in accuracy with the PBE correlation
over the whole periodic table of elements. This result can
be rationalized considering that in fact the two functionals
are constructed by imposing the same exact constraints.
Moreover, it supports the idea of using the jellium with gap
model as a reference system to describe the correlation energy
of inhomogeneous electron systems.

To complete our assessment we report in Table \ref{table3}
jellium surface correlation energies calculated with different methods
and compare them with accurate diffusion Monte Carlo (DMC) results \cite{PhysRevB.76.035403}.
\begin{table}
\begin{center}
\caption{\label{table3} Semi-infinite jellium surface correlation energies (erg/cm$^2$) for 
different values of the bulk parameter $r_s$ as computed with different functionals. Diffusion 
Monte Carlo (DMC) results \cite{PhysRevB.76.035403} are also given as reference. The last lines report the mean absolute error (MAE) and the mean absolute relative error (MARE). The best result for each 
line is highlighted in bold style. Results within the DMC error bar are underlined.}
\begin{ruledtabular}
\begin{tabular}{crrrrrr}
$r_s$ & LDA & PBE & PBEloc & GAPc & GAPloc & DMC \\
\hline
2 & 318 & 829 & \underline{\textbf{773}} & \underline{725} & 665 & 768 $\pm$50 \\
3 & 95 & 276 & \underline{\textbf{251}} & \underline{\textbf{233}} & 229 & 242 $\pm$10 \\
4 & 39 & 124 & \underline{112} & \underline{\textbf{103}} & \underline{\textbf{105}} & 104 $\pm$8 \\
6 & 10 & 40 & 35 & \underline{\textbf{31}} & 33 & 31 $\pm\ldots$ \\
  &   &    &    &     &  &  \\
MAE & 171 & 31 & \textbf{7} & 13 &  30 & \\
MARE & 62.4\% & 17.6\% & 6.2\% & \textbf{2.6\%} & 6.5\% & \\
\end{tabular}
\end{ruledtabular} 
\end{center}
\end{table}
The best results in this case are obtained by the
GAPc functional which yields for any $r_s$ results
in agreement with the reference DMC values and
the best MARE (it also gives the second best MAE; note 
anyway that the low MAE of PBEloc is mostly due to
its extremely good performance for $r_s=2$, while in the
other cases GAPloc has lower or equal absolute errors).
Thus, for the jellium surface correlation energies 
the GAPc functional definitely outperforms
the PBE functional. This traces back to the fact 
that the former recovers the correct second-order 
gradient expansion coefficient at any $r_s$ while 
the latter only in the high-density limit. In fact,
for GAPc the relative errors are almost constant at any
value of the bulk parameter,
while for PBE they increase significantly with $r_s$. 
Interestingly, the GAPc functional shows also a 
better overall performance than the PBEloc functional, 
which was parametrized on the jellium surface 
correlation energies.

Concerning the GAPloc functional we note that
it displays a remarkably good performance, being
even better than PBE (same MAE but better MARE)
and roughly comparable with the PBEloc
functional, especially at larger $r_s$ values 
(larger MAE but same MARE). This result
is an important achievement for the GAPloc functional
since it was constructed disregarding the
satisfaction of the second-order gradient
expansion and, even more importantly, because
for a quite wide range of $t$ values around zero
it recovers very closely LDA, which is
indeed very bad for this problem.
Nonetheless, we recall that the 
correct behavior in the slowly-varying limit
is not the only important feature in the
description of jellium surface energies, because this problem
also shows important contributions from the
rapidly-varying regions that are present outside the
surface. Thus, the crucial ingredient for the
accurate calculation of jellium surface energies is the
correct balance of the two limits.
This balance appears to be well described by the GAPloc
functional which thus yields reasonably accurate
results for jellium surface correlation energies. 

Finally, we recall that KCISK meta-GGA correlation functional 
improves over the KCIS meta-GGA correlation functional, for jellium surface correlation energies, but 
it is slightly worse than the PBE functional (see Fig. 6 and Table 4 of Ref. \citenum{kcisk}).

\subsection{Performance in combination with exchange}
To test further the capabilities of the correlation functionals
we considered their use with exchange.
To this end, at first we tested several combinations of our correlation
with existing GGA exchange functionals. We found that the
GAPc functional is rather well compatible with the revPBE
exchange \cite{revpbe}. On the other hand, no good exchange counterpart
was found for the GAPloc functional, probably because the
peculiar form of this functional requires ad hoc features in
the exchange part which are not included in present semilocal
functionals. Nevertheless, the GAPloc functional was found to
be well compatible with Hartree-Fock exchange. Thus,
we additionally considered a set of tests of our correlation 
functionals in combination with the exact Hartree-Fock exchange.
We remark that, since the focus 
of the present paper is on correlation functionals
based on the jellium with gap model, the testing within a XC
approach has the only scope to demonstrate the possible compatibility
of the present correlation methods in such a scheme, whereas the
development of an optimal semilocal exchange for the GAPc and especially the
GAPloc functionals is left for future work. At the same time the use
of the functionals in combination with Hartree-Fock exchange is
only intended to show the actual compatibility
of the methods with the exact exchange, without any intention
to solve more subtle problems related to static correlation or
non-local effects (see also Ref. \citenum{pbeloc}).

\begin{table}
\begin{center}
\caption{\label{tab4}Mean absolute errors (kcal/mol) on several tests for different 
exchange-correlation functionals. The label X+GAPc denotes the functional obtained composing the 
revPBE exchange with the GAPc correlation functional. The last line reports the average relative 
deviation with respect to PBE (ARD$_\mathrm{PBE}$; see Eq. (\ref{e43})). The best GGA result for 
each line is highlighted in bold style. The cases where X+GAPc is equal or better than revPBE are 
indicated with a star.}
\begin{ruledtabular}
\begin{tabular}{lrrrrrr}
Test & LDA & BLYP & PBE & APBE & revPBE & X+GAPc \\
\hline
AE6    & 75.6 & 6.8 & 14.2 & 7.8 & 9.0 & \textbf{6.2}$^*$ \\
W4     & 44.0 &	\textbf{5.8} & 10.7 & 8.6 & 6.7 & 6.9 \\
TM10AE & 29.7 &	13.8 & 12.8 & 11.5 & \textbf{10.7} & 12.2 \\
BH6    & 18.0 & 8.2 & 9.4 & 8.3 & 6.8 & \textbf{5.6}$^*$ \\
K9     & 14.6 & 6.0 & 7.3 & 6.5 & \textbf{4.9} & \textbf{4.9}$^*$ \\
DC9    & 156.9 & 26.5 & 40.8 & 29.4 & 34.2 & \textbf{24.9}$^*$ \\
G21IP  & 4.8 & 4.8 & 3.9 & 4.0 & 4.2 & \textbf{2.8}$^*$ \\
EA13   & 3.4 & 9.5 & 7.4 & 7.5 & 8.7 & \textbf{6.0}$^*$ \\
PA13   & 4.3 & 2.4 & \textbf{2.2} & 2.8 & 4.0 & 4.1 \\
AE17   & 426.5 & \textbf{7.4} & 51.6 & 22.2 & 13.6 & 11.6$^*$ \\
       &       &     &      &      &      & \\
ARD$_\mathrm{PBE}$ & 3.14 & 0.82 & 1.00 & 0.85 & 0.86 & \textbf{0.75}$^*$ \\
\end{tabular}
\end{ruledtabular}
\end{center}
\end{table}
In Table \ref{tab4} we report the mean absolute errors of several 
tests on atoms and molecules as resulting from different GGA approaches.
The functional obtained by the combination of revPBE exchange and GAPc 
correlation has been labeled X+GAPc. In the last line we report
the average relative deviation with respect to PBE (ARD$_\mathrm{PBE}$)
defined as
\begin{equation}\label{e43}
\mathrm{ARD}_\mathrm{PBE} = 
\frac{1}{8}\sum_{i=1}^8\frac{\mathrm{MAE}_i(\mathrm{method})}{\mathrm{MAE}_i(\mathrm{PBE})}\ ,
\end{equation}
where $\mathrm{MAE}_i(\mathrm{method})$ denotes the MAE of any given method for the $i$-th test.
This indicator provides a fair overall assessment of the whole set of tests
\cite{mukappa,bloc,kpbeint,hpbeint}.

We remark that the set of tests in Table \ref{tab4} does not provide, nor is intended
to provide, a throughout assessment of the functionals but aims instead
at demonstrating the compatibility of the GAPc functional with semilocal
exchange over a rather broad range of problems. This compatibility
is indeed well evident by inspecting the results reported in Table \ref{tab4}.
In fact, the X+GAPc functional shows a performance comparable with
that of the best GGA XC functionals (BLYP \cite{b88,lyp} and APBE \cite{apbe}).
More importantly, the comparison between revPBE and X+GAPc
results shows that the latter are in general superior to the former, 
despite the exchange part was not optimized for use with GAPc.
In particular, we remark the important improvement for ionization potentials (G21IP),
and electron affinities (EA13) since these properties may be closely 
related to opening of a gap in the the jellium with gap model system. 
This indicates the goodness of the GAPc correlation and suggest that
it can be a promising tool for future development of a
semilocal XC functional.

As additional test we report in Table \ref{tab5} the performance
of several GGA correlation functionals in combination with
Hartree-Fock exchange for a selected set of properties
(see also Ref \citenum{pbeloc}).
\begin{table}
\begin{center}
\caption{\label{tab5}Mean absolute errors (kcal/mol) on several tests as computed with 
Hartree-Fock exchange complemented with different GGA correlation functionals. The best value in 
each line is highlighted in bold style.}
\begin{ruledtabular}
\begin{tabular}{lrrrrr}
&\multicolumn{5}{c}{Hartree-Fock +}\\
\cline{2-6}
Test & LYP & PBE & PBEloc & GAPc & GAPloc \\
\hline
AE6 & 38.2 & 31.9 & 24.0 & 33.5 & \textbf{23.1} \\
BH6 & 5.3 & 5.6 & 4.4 & 6.1 & \textbf{3.7} \\
K9 & 6.0 & 5.7 & 4.7 & 6.0 & \textbf{3.4} \\
HB6 & 2.3 & 1.5 & 1.7 & 1.4 & \textbf{1.3}\\
\end{tabular}
\end{ruledtabular}
\end{center}
\end{table}
Inspection of the table shows that, as discussed above,
the GAPloc functional is definitely more
compatible with Hartree-Fock exchange that other GGA
functionals, outperforming for all the tests considered also the
PBEloc correlation \cite{pbeloc} which was constructed to
enhance such compatibility.
As we will show in next subsection, the good results of the GAPloc functional in this context
may be possibly traced back to its ability to describe with good accuracy the
correlation energy density of different systems
so that it can be summed to the Hartree-Fock energy density,
which is long-range, with small error accumulation (see also next subsection).
For this reason the functional can yield also very
accurate integrated correlation energies (see Table \ref{tab2}).
Possibly for the same reason the GAPloc functional is
hardly compatible with existing semilocal exchange approximations,
because these provide a too poor description of 
long-range effects (that in common XC functionals are
probably described by the correlation part). In fact,
in our tests the GAPloc correlation showed a reasonable performance
in combination with a PBE-like exchange functional \cite{pbe} with
enhancement factor $F_x(s)=1+\kappa-\kappa/(1+\mu s^2/\kappa)$
only for very large values of $\kappa$, i.e. when 
a high nonlocality is introduced into the functional \cite{mukappa}.
Finally, we remark the good performance of HF+GAPloc for the kinetics (K9 test),
being better than GGAs, many meta-GGAs \cite{bloc}, and hybrids functionals 
\cite{hpbeint}.

Finally, we mention that both KCIS and KCISK correlation 
functionals are well compatible with semilocal exchange. 
Thus, the PKZB meta-GGA exchange \cite{pkzb} combined 
with KCISK is accurate for atomization energies of small molecules 
\cite{kcisk}, while TPSS exchange \cite{tpss} combined 
with KCIS correlation gives good results for many properties \cite{zaho05}.

\subsection{Real-space analysis of GAPloc}
The results of Table \ref{tab5} show that GAPloc
has a higher compatibility with Hartree-Fock exchange.
We argued that this property depends on the good
shape of its correlation energy density.
In order to understand better this point we
perform in this subsection the real-space analysis \cite{parrbook,454274}
of the GAPloc correlation 
\begin{equation}\label{eee13}
E_c = 4\pi\sum_\sigma N_\sigma\int_0^\infty u^2\frac{\langle n_c\rangle_\sigma(u)}{2u}du\ ,
\end{equation}
where $\sigma$ is a spin index, $N_\sigma$ is the number of electrons with spin $\sigma$
and the angle- and system-averaged correlation hole is
\begin{equation}
\langle n_c\rangle_\sigma(u) = \frac{1}{N_\sigma}\int d\R 
n_\sigma(\R)\int\frac{d\Omega_u}{4\pi}\bar{n}_{c\sigma}(\R,\R+\mathbf{u})\ ,
\end{equation}
with $d\Omega_u$ the solid angle element in the $\mathbf{u}$-space.
The coupling-constant averaged correlation hole is constructed using
the reverse-engineering method of Ref. \citenum{blochole} (see the Appendix).
Note that the quantity $2\pi N_\sigma u\langle n_c\rangle_\sigma$ has the
dimension of a spherically-averaged energy density and its 
%
analysis allows to inspect the physical 
content of a given functional with high accuracy by comparison with
accurate reference benchmarks.
The correlation hole is in fact not only a space-resolved
expansion of the correlation energy, but unlike the correlation
energy density is uniquely defined \cite{PhysRevB.54.16533}. Moreover, we recall that,
at full coupling strength, the correlation hole is an observable \cite{PhysRevB.54.16533}.
We must stress anyway that while the PBE correlation hole was
constructed entirely from physical exact conditions, 
the reverse engineering method used in the construction of the GAPloc 
correlation hole is not unique, depending on the GAPloc 
correlation energy density. However, the here 
proposed GAPloc hole satisfies important exact conditions 
(e.g. hole sum rule, energy sum rule, accurate LDA on-top hole, 
and RPA non-oscillatory long-range contribution; see appendix and 
Ref. \citenum{blochole}), and thus it can be seen as a 
practical tool which reveals the physics behind the GAPloc correlation functional. 
We also recall that the TPSS and PBE correlation holes of Refs. 
\citenum{tpsshole,PhysRevB.54.16533,pitarke06} are very well mimicked by the reverse 
engineering hole method (see Fig. 2 of Ref. \citenum{blochole}) 
inside of the Coulomb hole radius.

\begin{figure}
\begin{center}
\includegraphics[width=\columnwidth]{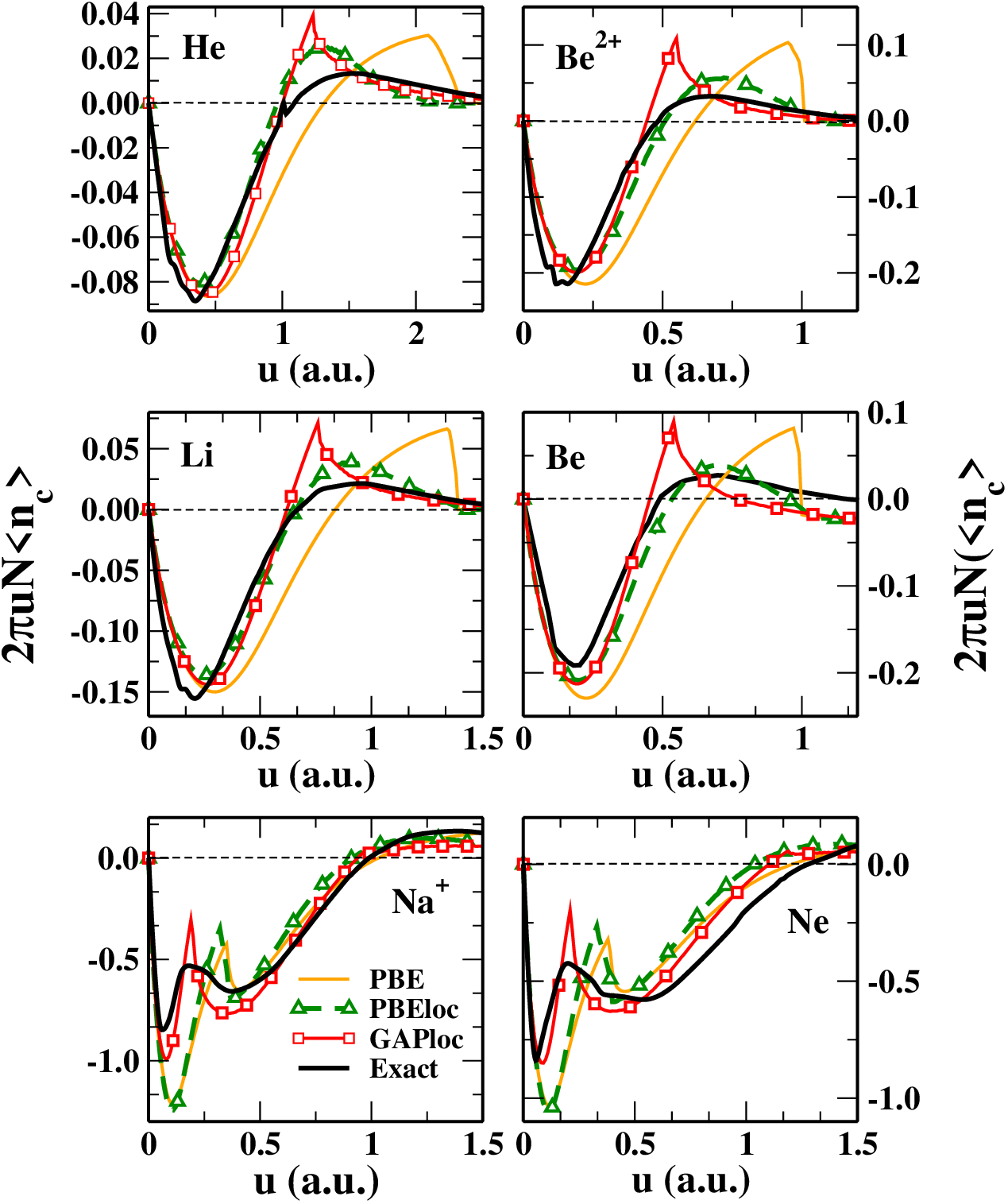}
\caption{\label{fig3}Real-space analysis of the correlation energy for several atoms. Reference 
data are taken from Refs. \citenum{0370-1328-78-5-328,477484}. Note that all the
curves in the plots integrate to the respective total correlation energy.}
\end{center}
\end{figure}
In Fig. \ref{fig3} we report the real-space analysis of the GAPloc correlation
energy for several atoms and ions. 
The PBE and PBEloc as well reference curves
are also reported for comparison.
The plot shows that overall the GAPloc functional provides
a better description of the correlation hole, in
particular at short range which is the most
relevant for any semilocal functional
(whereas the long-range part can be hardy described accurately
at this level of theory).
On the other hand, at longer range the GAPloc plots, similarly to
PBE, show some cusp features due to the
cutoff procedure used to ensure the proper
normalization of the hole during its construction
\cite{blochole,PhysRevB.54.16533} 
(see Eq. (\ref{ae13})).
On the contrary PBEloc displays smoother hole curves because, 
thanks to its very rapid decay in the 
tail of the density, the cuttoff procedure is better averaged 
in the process of real-space calculation.

One more important observation is that
the GAPloc functional describes reasonably well 
also the larger atoms, i.e. Na$^+$ and Ne, giving
a good and balanced description of all the features of their 
correlation hole. This is a remarkable achievement of
the GAPloc functional, since other GGA functionals,
and even meta-GGA ones \cite{blochole}, perform rather worse
in this context.

To make the analysis more quantitative we report in
Table \ref{tab6} the Coulomb hole radius as
obtained from the different GGA holes.
We recall that the Coulomb hole radius is the 
smallest distance ($u\neq 0$) where the correlation 
hole equals zero. This is an important quantity
to define the effect of correlation on the distribution
of the electrons in the vicinity of each other \cite{blochole,PhysRevA.51.2815}.
\begin{table}
\begin{center}
\caption{\label{tab6}Coulomb hole radius as obtained from different GGA approximations for several 
ions and atoms. The last lines report the mean absolute error (MAE) and the mean absolute relative 
error (MARE). The best result in each line is denoted with bold style. Reference values are taken 
from Refs. \cite{PhysRevA.68.022505,0370-1328-78-5-328,477484,2033747}}
\begin{ruledtabular}
\begin{tabular}{lccccc}
Atom & LDA & PBE & PBEloc & GAPloc & Ref. \\
\hline
He & 3.51 & 1.32 & \textbf{0.97} & 1.00 & 0.93 \\
Li$^+$ & 2.53 & 0.84 & \textbf{0.67} & 0.63 & 0.67 \\
Be$^{2+}$ & 2.06 & 0.62 & \textbf{0.51} & 0.45 & 0.49 \\
B$^{3+}$ & 1.77 & 0.49 & 0.42 & \textbf{0.35} & 0.35 \\
Ne$^{8+}$ & 1.15 & 0.25 & 0.23 & \textbf{0.16} & 0.17 \\
Ca$^{18+}$ & 0.78 & 0.13 & 0.12 & \textbf{0.08} & 0.08 \\
Li & 2.53 & 0.85 & \textbf{0.67} & 0.63 & 0.67 \\
Be & 5.23 & 0.65 & 0.54 & \textbf{0.46} & 0.48 \\
Ne & 2.35 & \textbf{1.21} & 1.04 & 1.11 & 1.27 \\
Na$^{+}$ & 2.05 & 1.10 & 0.92 & \textbf{0.94} & 0.98 \\
        &      &       &      & & \\
MAE & 1.79 & 0.15 & 0.06 & \textbf{0.04} & \\
MARE & 420\% & 32\% & 15\% & 5\% & \\
\end{tabular}
\end{ruledtabular}
\end{center}
\end{table}
The data in the table confirm that the GAPloc functional
is the most accurate for the description 
of the Coulomb hole, with a MAE of 0.04 Bohr
and a MARE of 5\%. Note that this is a remarkable performance,
even outperforming meta-GGA functionals \cite{blochole}.
Good results are found also with the PBEloc
correlation functional which yields a MAE of 
0.06 Bohr, similar to the BLOC meta-GGA correlation functional \cite{bloc,blochole}.
Note that the PBE functional displays also a good 
performance, greatly improving over the LDA, but still shows 
a systematic overestimation of the
Coulomb hole radius.

\section{Conclusions}
In this paper we have studied the jellium with gap model in the context of the 
ground-state density functional theory. First, we have proposed a reparametrization of 
the spin-dependence of the jellium with gap correlation energy (in the spirit of the 
Perdew-Wang LDA correlation parametrization, see Table \ref{tab1}). 
Next, we derived exact constraints for 
the local gap function, starting from known exact properties of the correlation energy. 
This analysis allowed us to construct two semilocal correlation functionals,
namely GAPc and GAPloc, that keep the original jellium with gap functional
form and are characterized solely by the modelling of the local gap function $G[n,\nabla n]$.

The GAPc functional recovers the correct second-order gradient expansion at any density, 
satisfies various density-scaling properties, and has a slower decay in the tail of the density than 
the PBE functional (see Fig. \ref{fig1}). It is well compatible with the revPBE exchange functional, 
which is considered one of the most accurate semilocal exchange, improving the total MAE 
of various properties and test sets with about 15\% (see Table \ref{tab4}). It 
competes with state-of-the-art
GGA functionals for the correlation energies of atoms and ions (see Table \ref{tab2}), while 
outperforms them for the jellium surface correlation energies (see 
Table \ref{table3}), suggesting that it can be a useful tool for solid-state physics
if combined with a proper semilocal exchange.

The GAPloc functional satisfies various density-scaling properties, and recovers the LDA 
functional over a wide range of slowly-varying regimes. This later feature is a direct consequence 
of the fitting to the exact correlation energy density of He atom. It is very accurate for total 
correlation energies of atoms and ions (see Table \ref{tab2}), performs
reasonably well for jellium surface correlation energies (see Table \ref{table3}), 
and is one of 
the most compatible GGA correlation functionals with the exact exchange (see
Table \ref{tab5}).  
Moreover, by constructing its underlying correlation hole, we have shown that it gives 
the most realistic real-space analysis of the atomic correlation energy (see
Fig. \ref{fig3}), 
and Coulomb hole radii (see Table \ref{tab6}). These results 
suggest that GAPloc can be also used 
in construction of more sophisticated functionals (e.g. hybrids and hyper-GGAs). 

Finally, we mention that this work provides indications that the second-order gradient expansion of 
the correlation energy introduces a gauge in the correlation energy density which is not 
compatible with the exact exchange, but it is with semilocal exchange. Thus, it is responsable for 
an important part of the exchange-correlation error compensation.

\appendix

\section{Local density approximation with a gap and its asymptotic properties}
\label{appa}
The main equation defining the LDA with gap model is Eq. (\ref{ee4}).
This uses as main ingredients the convetional LDA correlation energy per
particle, here in the Perdew-Wang parametrization \cite{pwlda},
\begin{eqnarray}
\epsilon_c^{LDA}(r_s) & = & -2A(1+\alpha r_s)\times\\
\nonumber
&& \times \log\Big[1+\frac{1}{2A\left(\beta_1r_s^{1/2}+\beta_2r_s+\beta_3r_s^{3/2}+\beta_4r_s^2\right)}\Big]\ ,
\end{eqnarray}
and the functions $c_1$, $c_2$, and $c_3$ are  
\begin{equation}
c_1(r_s) = C\frac{2\left(\epsilon'\right)^2-\epsilon_c^{LDA}\epsilon''}
{2\left(C\epsilon'-\left(\epsilon_c^{LDA}\right)^2\right)}
\end{equation}
\begin{equation}
c_2(r_s) = \frac{2\epsilon_c^{LDA}\epsilon' - C\epsilon''}
{2\left(C\epsilon'-\left(\epsilon_c^{LDA}\right)^2\right)}
\end{equation}
\begin{equation}
c_3(r_s) = -\frac{2\left(\epsilon'\right)^2-\epsilon_c^{LDA}\epsilon''}
{2\left(C\epsilon'-\left(\epsilon_c^{LDA}\right)^2\right)}\ ,
\end{equation}
where
\begin{equation}
C(r_s) = f_cr_s^{-2}
\end{equation}
\begin{equation}
\epsilon'(r_s) = \frac{a_1r_s^{3/2}}{1+a_2r_s^{1/2}+a_3r_s+a_1r_s^{3/2}}
\end{equation}
\begin{equation}
\epsilon''(r_s) = \sum_{i=3}^7b_ir_s^i\ .
\end{equation}
The numerical values of all the parameters, for both the spin-unpolarized and
the fully-spin-polarized case, are reported in Table \ref{tab1}.
Note that with respect to Ref. \citenum{kcisk} for the  fully-spin-polarized case
we performed a reparametrization of $\epsilon'$, $\epsilon''$, and
$C$ in order to remove the prefactors from the definition of 
$c_1$, $c_2$, $c_3$. This makes the formulas for $\epsilon_0$ and
$\epsilon_1$ formally identical and simplifies the notation.
Our fit agrees with the results of Ref. \citenum{kcisk} within 0.1 mHa
(note that this is one order of magnitude smaller than the
accuracy of the original fit).
\begin{table}
\begin{center}
\caption{\label{tab1}Parameters defining the jellium with gap model}
\begin{ruledtabular}
\begin{tabular}{lll}
\multicolumn{3}{c}{$\zeta=0$} \\
$A=0.031091$ & $b_3=-2.504\cdot 10^{-2}$ & $a_1= 0.004953$ \\
$\alpha=0.21370$ & $b_4=7.026\cdot 10^{-3}$ & $a_2=1.07024$ \\
$\beta_1=7.5957$ & $b_5=-1.268\cdot 10^{-3}$  & $a_3=0.07928$ \\
$\beta_2=3.5876$ & $b_6=1.136\cdot 10^{-4}$ & \\
$\beta_3=1.6382$ & $b_7=-3.841d\cdot 10^{-6}$ & $f_c=0.23878$ \\
$\beta_4=0.49294$ & & \\
\hline
\multicolumn{3}{c}{$\zeta=1$} \\
$A=0.015545$ & $b_3=-3.24091\cdot 10^{-2}$ & $a_1=0.0471985$ \\
$\alpha=0.20548$ & $b_4=9.99978\cdot 10^{-3}$ &  $a_2=1.49676$ \\
$\beta_1=14.1189$ & $b_5=-1.93483\cdot 10^{-3}$ &  $a_3=0.00179054$ \\
$\beta_2=6.1977$ & $b_6=1.79118\cdot 10^{-4}$ &  \\
$\beta_3=3.3662$ & $b_7=-6.15798\cdot 10^{-6}$ &  $f_c=0.0645351$ \\
$\beta_4=0.62517$ & & \\
\end{tabular}
\end{ruledtabular}
\end{center}
\end{table}

In the low-density limit ($n\rightarrow0$ and
$r_s\rightarrow\infty$) the functions $c_1$, $c_2$, 
and $c_3$ behave as
\begin{equation}
c_1 \propto \frac{b_7}{2\beta_4}r_s^6 \rightarrow -\infty
\end{equation}
\begin{equation}
c_2 \propto -\frac{b_7}{2}r_s^7 \rightarrow +\infty
\end{equation}
\begin{equation}
c_3 \propto -\frac{b_7}{2\beta_4}r_s^8 \rightarrow +\infty\ .
\end{equation}
Thus, we have for the spin-polarized and -unpolarized 
correlation energies per particle 
\begin{equation}\label{e15}
\epsilon(r_s,G) \propto 
\frac{\left(-\frac{1}{\beta_4r_s}\right)+\frac{b_7}{2\beta_4}r_s^6G}
{1-\frac{b_7}{2}r_s^7G-\frac{b_7}{2\beta_4}r_s^8G^2}\ .
\end{equation}

On the other hand, in the high-density limit, when
$n\rightarrow\infty$ and $r_s\rightarrow0$ we have
\begin{eqnarray}
c_1 & \propto & -\frac{Ab_3}{2a_1}r_s^{3/2}\log(r_s)\rightarrow 0^-\\
c_2 & \propto & -\frac{b_3}{a_1}r_s^{3/2} \rightarrow 0^+ \\
c_3 & \propto &  \frac{Ab_3}{2f_ca_1}r_s^{7/2}\log(r_s) \rightarrow 0^+\ .
\end{eqnarray}
Therefore, the spin-polarized and -unpolarized 
correlation energies per particle behave as
\begin{equation}\label{e19}
\epsilon(r_s,G) \propto \frac{A\log(r_s) 
-A\frac{b_3}{2a_1}r_s^{3/2}\log(r_s)G}{1-\frac{b_3}{a_1}r_s^{3/2}G + 
\frac{Ab_3}{2f_ca_1}r_s^{7/2}\log(r_s)G^2}\ .
\end{equation}

\subsection{Exact constraints for the local gap function}
Several exact constraints are known for the correlation energy and
can be used to construct accurate approximate correlation functionals
with minimal empiricism. In this subsection we investigate how
these constraints apply to the LDA model with a gap and,
in particular, we consider the corresponding requirements for a 
local gap function $G(r_s,t)$.

\subsubsection{Uniform electron gas and slowly-varying density limits}
In the limit of the uniform electron gas ($\nabla n=0$) we must have
$\epsilon_c=\epsilon_c^{LDA}$. Thus, we must require that the gap
vanishes wherever $\nabla n=0$, i.e. we must impose that
$G\propto t^\gamma$ with $\gamma>0$.

In the slowly-varying density limit ($t\rightarrow 0$) the
correlation energy per particle is described by
the second-order gradient expansion \cite{pbe,PhysRevB.43.8911}
\begin{equation}
\epsilon_c \approx \epsilon_c^{LDA} + \phi^3\beta(r_s)t^2\ ,
\end{equation}
with $\beta$ the (eventually $r_s$-dependent) second-order gradient
expansion correlation coefficient \cite{PhysRev.165.18,PhysRevB.33.943}.
To fulfill the second-order gradient expansion any functional must satisfy
the condition
\begin{equation}
\phi^3\beta(r_s) = \frac{\partial\epsilon_c}{\partial t^2}\Big|_{|\nabla n|^2=0}\ .
\end{equation}
Using the chain rule
\begin{equation}
\frac{\partial\epsilon_i}{\partial t^2}\Big|_{|\nabla n|^2=0} = \frac{\partial\epsilon_i}{\partial 
G}\Big|_{G=0}\frac{\partial G}{\partial t^2}\Big|_{t^2=0}\ ,
\end{equation}
we thus find
\begin{eqnarray}
\nonumber
\phi^3\beta(r_s) & = & \frac{\partial\epsilon_0}{\partial G}\Big|_{G=0}\frac{\partial G(\zeta=0)}{\partial t^2}\Big|_{t^2=0}\left(1-f(\zeta)\right) + \\
&& + \frac{\partial\epsilon_1}{\partial G}\Big|_{G=0}\frac{\partial G(\zeta=1)}{\partial t^2}\Big|_{t^2=0}f(\zeta)\ .
\end{eqnarray}
This condition cannot be easily satisfied for any $\zeta$, without a strong
modification of Eq. (\ref{e2}). Nevertheless, it can be easily satisfied
in both the spin-unpolarized and the full-spin-polarized limits,
yielding the condition
\begin{equation}
\label{e24}
\frac{\partial G}{\partial t^2}\Big|_{t^2=0} = 
\phi^3\beta(r_s)\left[\frac{\partial\epsilon}{\partial G}\Big|_{G=0}\right]^{-1} 
=\phi^3\frac{\beta(r_s)}{c_1-c_2\epsilon_c^{LDA}}\ . 
\end{equation}

The uniform electron gas limit and Eq. (\ref{e24}) are fulfilled by a local
gap function of the general form
\begin{equation}\label{e25}
G(r_s,t^2,\zeta) = \phi^3 \frac{\beta(r_s)t^2}{c_1-c_2\epsilon_c^{LDA}}H(r_s,t^2)\ ,
\end{equation}
where $H(r_s,0)=1$ and $[t^2(\partial H/\partial t^2)]_{t^2=0}=0$.

\subsubsection{Rapidly-varying density limit}
In the rapidly-varying density limit ($t\rightarrow\infty$) the correlation
energy must vanish \cite{pbe}. Therefore, we must impose that
in this limit $G\rightarrow \infty$. For the local gap function
of Eq. (\ref{e25}) this implies that we must have
$H(r_s,t^2\rightarrow\infty)\propto t^{2\gamma}$ with $\gamma > -1$.

\subsubsection{Uniform scaling to the high-density limit}
Under the uniform scaling to the high-density limit \cite{PhysRevA.45.1509}
$n(\R)\rightarrow \lambda^3n(\lambda\R)$ 
with $\lambda\rightarrow\infty$. 
Thus, $r_s\rightarrow 0$ as $\lambda^{-1}$ 
and $t^2\rightarrow \infty$ as $\lambda$.
In this limit the correlation energy per particle
must scale to a constant \cite{pbe,QUA:QUA560360864}.
However, the LDA correlation is diverging logarithmically \cite{pwlda}.

According to Eq. (\ref{e19}), under the uniform 
scaling to the high-density limit,
the fully-spin-polarized and the spin-unpolarized 
correlation energies per particle of the jellium with gap model
behave as
\begin{equation}\label{e26}
\epsilon \rightarrow \frac{A\log(\lambda^{-1}) 
-A\frac{b_3}{2a_1}\lambda^{-3/2}\log(\lambda^{-1})G}{1-\frac{b_3}{a_1}\lambda^{-3/2}G + 
\frac{Ab_3}{2f_ca_1}\lambda^{-7/2}\log(\lambda^{-1})G^2}\ .
\end{equation}
To cancel the logarithmic divergence of the LDA correlation
we must therefore require that $G\rightarrow (2a_1/b_3)\lambda^{3/2}$
or $G$ diverges faster than $\lambda^{7/4}$.

In the first case, for the local gap function of Eq. (\ref{e25}) we must have
\begin{equation}
\frac{\beta(0)\lambda^{5/2}\log^{-1}(\lambda^{-1})}{A\frac{b_3}{2a_1}}H = 
\frac{2a_1}{b_3}\lambda^{3/2}\ .
\end{equation}
This condition is satisfied by
\begin{equation}
H(r_s\rightarrow 0,t^2\rightarrow\infty) \propto \frac{Ar_s\log(r_s)}{\beta(0)}\ .
\end{equation}

\subsubsection{Thomas-Fermi scaling}
The Thomas-Fermi scaling \cite{PhysRevLett.100.256406,scaling} 
is defined by the transformation
$n(\R)\rightarrow\lambda^2n(\lambda^{1/3}\R)$ with $\lambda\rightarrow\infty$.
Hence, $r_s\rightarrow0$ as $\lambda^{-2/3}$, while $t$ is unchanged.
Under this scaling the semiclassical high-density limit is reached.
In this condition the correlation is dominated by the LDA 
contribution \cite{scaling}, thus it must scale as $A\log(\lambda^{-2/3})$.

When the Thomas-Fermi scaling is set up, 
according to Eq. (\ref{e19}), the fully-spin-polarized and the spin-unpolarized 
correlation energies per particle of the jellium with gap model
behave as
\begin{equation}
\epsilon \rightarrow \frac{A\log(\lambda^{-2/3}) 
-A\frac{b_3}{2a_1}\lambda^{-1}\log(\lambda^{-2/3})G}{1-\frac{b_3}{a_1}\lambda^{-1}G + 
\frac{Ab_3}{2f_ca_1}\lambda^{-7/3}\log(\lambda^{-2/3})G^2}\ .
\end{equation}
Thus, to achieve the proper scaling the local gap function must 
behave in this limit as $G\propto \lambda^\gamma$ with $\gamma<1$.

For the local gap function of Eq. (\ref{e25}) this condition 
implies
\begin{equation}
\lambda\log^{-1}(\lambda^{-2/3})H\propto \lambda^\gamma\; \textrm{with}\; \gamma<1\ .
\end{equation}
Hence, we must have that $H$ 
is not diverging faster than $\log(\lambda^{-2/3})$.

\subsubsection{Tail behavior}
Although there are no exact constraints known for the
behavior of the correlation energy per particle in the tail
of atomic systems it is interesting to investigate
the decay of correlations functionals in this situation.
In fact, recent work highlighted the importance of the
tail behavior for approximate correlation functionals \cite{pbeloc}.

In the tail of an atom the electron density has the asymptotic
form \cite{PhysRevA.30.2745}
\begin{equation}
n(r) \propto e^{-2\sqrt{-2\epsilon_H}r}\; \mathrm{for}\; r\rightarrow\infty\ ,
\end{equation}
with $\epsilon_H$ the energy of the highest occupied orbital.
Thus, of course $r_s\rightarrow\infty$.
The reduced gradient for correlation is consequently
\begin{equation}
t^2\propto \frac{-\epsilon_He^{-4\sqrt{-2\epsilon_H }r}}{e^{-(14/3)\sqrt{-2\epsilon_H }r}} \sim 
-\epsilon_H r_s\ .
\end{equation}
Therefore, both the Seitz radius and the square of the reduced gradient for
correlation show the same decay behavior in the tail of an
atom. We formalize this situation by introducing a 
scaling parameter $\lambda\rightarrow\infty$ such that
$r_s\propto\lambda$ and $t^2\propto\lambda$.

The behavior of the fully-polarized and unpolarized correlation
energies per particle in this regime is described
by Eq. (\ref{e15}). We see that for any local
gap function not vanishing faster than $\lambda^{-7}$
the decay behavior of the correlation energies per particle is
\begin{equation}
\epsilon(r_s,G) \propto -\frac{1}{\lambda^2G}\ .
\end{equation}
Note that in contrast the decay behavior of the
PBE correlation functional is \cite{pbeloc}
\begin{equation}
\epsilon_c^{PBE}\propto \frac{Q}{\beta^2t^4}\ ,
\end{equation}
where
\begin{equation}
Q =\gamma^3\phi^3\left[e^{\epsilon_c^{LDA}/(\gamma\phi^3)}-1\right]^3e^
{-2\epsilon_c^{LDA}/(\gamma\phi^3)}\ ,
\end{equation}
with $\gamma$ a constant. Thus, in the tail of an atomic density we have
\begin{equation}
\epsilon_c^{PBE} \propto -\frac{1}{\lambda^5}\ .
\end{equation}

\section{GGA hole model}
In Ref. \citenum{blochole} the reverse engineering
hole model was shown for the most general case of
meta-GGA functionals. Here we report explicit
formulas for the GGA case.
The spin- and angle-averaged correlation hole model is
\begin{eqnarray}
\label{ae13}
\bar{n}_c^{GGA}(\R,u) & = & \bar{n}_c^{GGA}[r_s(\R),\zeta(\R),t(\R)](v) = \\
\nonumber
& = & \phi^5k_s^2\Big[A_c(r_s,\zeta,v)+ t^2B_c^{GGA}(r_s,\zeta,t,v)\Big]\theta(v_c-v)\ ,
\end{eqnarray}
where
$v=\phi k_s u$ is the reduced electron-electron separation on the scale
of the screening length.
The function 
$\phi^5k_s^2A_c$ is the LDA correlation hole \cite{PhysRevB.54.16533,PhysRevB.46.12947}.
The function $B_c$ is chosen to be
\begin{eqnarray}\label{ae14}
B_c^{GGA}(r_s,\zeta,t,v) & = & B_c^{LM}(v)\left[1-e^{-\eta^{3}}\right] + \\
\nonumber
&& + \mu^{GGA}(r_s,\zeta,t)v^2e^{-\eta^{3/2}}\ .
\end{eqnarray}
Here $B_c^{LM}(v)$ is the RPA nonoscilating long-range contribution
\cite{PhysRevB.54.16533}, $\eta=\sqrt{p}v$ is
a scaled distance suitable for the gradient correction,
with $p(r_s,\zeta)=\pi k_F(0.305-0.136\zeta^2)/4\phi^4$ measuring where
the short range contribution vanishes.
The function $\mu^{GGA}$ is fixed by imposing the energy sum rule
$2\pi \int \bar{n}_c^{GGA}(\R,u)u du=\epsilon_c^{GGA}(\R)$. It is
\begin{eqnarray}
\label{ae15}
\mu^{GGA}(r_s,\zeta,t) & = & \Bigg[\frac{\phi^2k_s^2}{2\pi}\epsilon_c^{GGA} - k_s^2\phi^5\int_0^{v_c}A_cvdv -\\
\nonumber
&& - k_s^2\phi^5t^2\int_0^{v_c}B_c^{LM}(1-e^{-\eta^3})vdv\Bigg]\Bigg/\\
\nonumber
&& / \Big[k_s^2\phi^5t^2\int_0^{v_c}e^{-\eta^{3/2}}v^3dv\Big]\ .
\end{eqnarray}
This function controls the short-range (small $v$) behavior
of the hole model, which is the most important for a semilocal hole.
In contrast with the case of Ref. \citenum{PhysRevB.54.16533},
it is not constructed from the slowly-varying behavior of any underlying functional,
but it is instead entirely determined by the energy sum rule. Therefore,
it is more general and can be used to any semilocal correlation functional.


\end{document}